\documentstyle[12pt]{article}

\newlength{\dinwidth}
\newlength{\dinmargin}
\begin{document}

\def\bold#1{\setbox0=\hbox{$#1$}%
     \kern-.025em\copy0\kern-\wd0
     \kern.05em\copy0\kern-\wd0
     \kern-.025em\raise.0433em\box0 }
\def\slash#1{\setbox0=\hbox{$#1$}#1\hskip-\wd0\dimen0=5pt\advance
       \dimen0 by-\ht0\advance\dimen0 by\dp0\lower0.5\dimen0\hbox
         to\wd0{\hss\sl/\/\hss}}
\def\lq{\left [}
\def\rq{\right ]}
\def\AA{{\cal A}}
\def\HH{{\cal H}}
\def\LL{{\cal L}}
\def\MM{{\cal M}}
\def\VV{{\cal V}}

\newcommand{\be}{\begin{equation}}
\newcommand{\ee}{\end{equation}}
\newcommand{\bea}{\begin{eqnarray}}
\newcommand{\eea}{\end{eqnarray}}
\newcommand{\nn}{\nonumber}
\newcommand{\dd}{\displaystyle}
\newcommand{\bra}[1]{\left\langle #1 \right|}
\newcommand{\ket}[1]{\left| #1 \right\rangle}
\thispagestyle{empty}
\vspace*{1cm}
\rightline{BARI-TH/93-150}
\rightline{SISSA Ref. 114/93/EP}
\rightline{UTS-DFT-93-19}
\rightline{July 1993}
\vspace*{2cm}
\begin{center}
  \begin{Large}
  \begin{bf}
Radiative $B \to K^* \gamma $ transition in QCD \\
  \end{bf}
  \end{Large}
  \vspace{8mm}
  \begin{large}
P. Colangelo $^{a,}$ \footnote{e-mail address:
COLANGELO@BARI.INFN.IT},  C.A. Dominguez $^{c,}$ \footnote{On sabbatical
leave from Department of Physics, University of Cape Town, South Africa.},
 G. Nardulli $^{a,b}$
and N. Paver $^d$ \\
  \end{large}
  \vspace{6mm}
$^{a}$ Istituto Nazionale di Fisica Nucleare, Sezione di Bari, Italy\\
  \vspace{2mm}
$^{b}$ Dipartimento di Fisica, Universit\'a
di Bari, Italy \\
  \vspace{2mm}
$^{c}$ Scuola Internazionale Superiore di Studi Avanzati, Trieste,Italy \\
  \vspace{2mm}
$^{d}$ Dipartimento di Fisica Teorica, Universit\'a
di Trieste, Italy,  and \\
  \vspace{2mm}
Istituto Nazionale di Fisica Nucleare, Sezione di Trieste, Italy\\
  \vspace{2mm}

\end{center}
\begin{quotation}
\vspace*{1.5cm}
\begin{center}
  \begin{bf}
  ABSTRACT
  \end{bf}
\end{center}
\vspace*{0.5cm}

We evaluate the branching ratios of the
radiative decays $B \to K^*  \gamma $, and $B \to K_1 \gamma $
in the framework of three-point function QCD sum rules. We find:
$ \Gamma(B \to K^* \gamma) / \Gamma(b \to s \gamma) = 0.17 \pm 0.05$ and
$ \Gamma(B \to K_1 \gamma) / \Gamma(b \to s \gamma) = 0.30 \pm 0.15$.
We also study, in  the infinite quark mass limit, the
 connection between the form factor of the
radiative $B \to K^*  \gamma $ transition and
the weak hadronic form factors entering
the semileptonic B-meson decays.

\noindent
\end{quotation}

\newpage
\baselineskip=18pt
\setcounter{page}{1}

Radiative $B$-meson decays of the type $B \to K^*_i \gamma$, with
$K^*_i=K^*(892), \; K_1(1400)$, etc., are known to provide valuable information
on the Standard Model (SM) at the quantum level
 \cite{Review}. In fact,
these decays, induced by flavour changing
$b \to s $ neutral currents, are
controlled by the one-loop electromagnetic penguin operator which involves
important SM parameters such as the top quark mass and the
Cabibbo-Kobayashi-Maskawa matrix elements $V_{ts}$ and $V_{tb}$. At the quark
level, it has been shown \cite{MASIERO} that QCD corrections lift the
Glashow-Iliopoulos-Maiani suppression leading to a sizeable enhancement of the
branching ratio $BR(b \to s \gamma)$. Indeed, preliminary experimental results
have been recently reported from the CLEO II Collaboration
\cite{CLEO}:
\be BR(B \to K^*(892) \; \gamma) = (4.5 \pm 1.5 \pm 0.9) \times 10^{-5} \;.
\label{eq0}\ee

On the theoretical side, the calculation of the branching ratios of the
exclusive processes
induced by the $b \to s $ transition is
plagued by the usual uncertainties involved in determining
weak hadronic form factors. Estimates are
available e.g. from quark models \cite{ALTOMARI},
and from the Heavy Quark Effective Theory in the two extreme cases
where the chiral symmetry is assumed for the
light quark sector \cite{CASALBUONI}, or where the strange quark is considered
to be heavy \cite{ALI}; the different estimates
differ by up to an order of magnitude.

Among the various theoretical approaches, QCD sum rules
\cite{SVZ} appear to be promising since they are,
in principle, model independent. A calculation,
in this framework, of $B \to K^*_i \gamma$
has been done some time ago \cite{PAVER} using the formalism of two-point
functions. However, since $B^*_s$ cannot decay into $B \; K^*$ because of phase
space, two-point function QCD sum rules provide information only on the
coupling of the effective current to $B^*_s$. Vector Meson Dominance
must then be invoked, and an independent estimate of the strong
coupling constant $g_{B^*_s B K^*}$ is required.
The latter lies outside the framework of
two-point function QCD sum rules and, therefore, it is model dependent.
A possible way
out of this model dependency is offered by the formalism of QCD sum rules
for three-point functions. In this case the hadronic form factors entering
the matrix element of $B \to K^*_i \gamma$  can be determined provided the
leptonic decay constants of the $B$ and $K^*_i$ mesons are known.The former
can be estimated from two-point function QCD sum rules
 \cite{fb},
as well as
from lattice QCD \cite{Martinelli}, whereas the leptonic decay constant of
$K^*_i$ can be extracted e.g.
from  the $\tau$-lepton decay data \cite{Kolanoski}.

In this paper we pursue this three-point function QCD sum rule approach,
which is similar in spirit
to another calculation that can be found in the
literature \cite{ALIEV}; however, as it
will be shown below, the numerical results of \cite{ALIEV}
can be made more reliable,
and a more accurate determination of the above mentioned
decay channels can be obtained.In addition, we study here the connection
which follows in the $m_{b}\to \infty$ limit between the form factor of
$B\to K^{*} \gamma$ and the semileptonic B-meson form factors.

The electromagnetic penguin operator which governs the
$b \to s \gamma$ transition can be written,
for $m_s << m_b$, as follows:
\be \HH_{eff} = C m_b \epsilon^\mu {\bar s} \sigma_{\mu \nu}
{(1 + \gamma_5) \over 2} q^\nu b \hskip 3 pt , \label{eq1}\ee
\noindent
where $\epsilon$ and $q$ are the photon polarization and momentum,
respectively.
The constant $C$ contains the dependence on the Cabibbo-Kobayashi-Maskawa
angles and on the heavy quark masses; neglecting $m_c$ it reads:
\be C = {G_F \over \sqrt 2} {e \over 4 \pi^2} \; V^*_{ts} \; V_{tb} \;
F_2( { {m_t^2} \over {m_W^2}}) \hskip 5pt ,\label{eq2}\ee

\noindent where the function $F_2$ is given by \cite{MASIERO}:
\be F_2(x) = \eta^{-16/23} \Big \{ \bar F_2(x) + {116 \over 27} \Big [
{\eta ^{10/23} - 1 \over 5} + { \eta ^{28/23} - 1 \over 14} \Big] \Big \}
\hskip 5pt . \label{eq3} \ee

\noindent In eq.(\ref{eq3}) $\eta= \alpha_s(\mu) / \alpha_s(m_W)$ where
 $\mu \simeq m_b$ is the typical scale of the process, and
 $\bar F_2(x) $ reads \cite{INAMI} :
\be \bar F_2(x) = { x \over (x-1)^3} \; \Big[ {2 x^2 \over 3}  +
{5 x \over 12} - {7  \over 12} -
{3 x^2 - 2 x  \over 2 ( x-1) } \ln(x) \Big ] \hskip 5pt. \label{eq4} \ee

\noindent
The function $F_2(x)$ depends weakly on the top quark mass, e.g. in the range
$90\; GeV < m_t <  210 \; GeV$, $F_2(x)$ increases
from $0.55$ to $0.68$.
Using $m_b=4.6 \;GeV$, $\tau_B=1.4 \; ps$ and $m_t=120 \; GeV$, this
leads to the prediction
of the inclusive $b \to s \gamma$ branching ratio:
$BR(b \to s \gamma) = 2.2 \; (|V_{ts}|/0.042)^2 \times 10^{-4}$.

\par
Let us now consider the decay $B \to K^*(892) \; \gamma$.
According to eq.(\ref{eq1}) the amplitude for
$B(p) \to K^*(p',\eta) \; \gamma(q,\epsilon)$
can be written as follows:
\be \AA(B \to K^* \gamma) = \bra{K^*(p',\eta)} J^\mu_{eff} \ket {B(p)} \hskip
3pt \epsilon_\mu
\hskip 5pt ,\label{eq5}\ee

\noindent
where the effective current $J^\mu_{eff}$ is given by:
\be J^\mu_{eff} = C m_b \bar s  \sigma^{\mu \nu} {(1 + \gamma_5)\over 2}
q_\nu b \hskip 5pt. \label{eq6}
\ee

\noindent
 The
two hadronic form factors $F_1$ and $G_2$ appearing in the matrix element in
eq.(\ref{eq5}), i.e.
\bea \bra{K^*(p',\eta)} \bar s \sigma_{\mu \nu} {(1 + \gamma_5) \over 2}
q^\nu b \ket {B(p)}
& = &i \epsilon_{\mu \nu \rho \sigma} \eta^{*\nu} p^\rho p^{\prime \sigma}
F_1(q^2)
\nonumber \\
&+& [\eta^*_\mu (m^2_B - m^2_{K^*}) - (\eta^* \cdot q) (p + p^\prime)_\mu]
G_2(q^2)
 \label{eq7}\eea

\noindent ($q=p-p'$)
can be related
using the identity
$\sigma^{\mu \nu} \gamma_5= - {i \over 2} \epsilon^{\mu \nu \rho \sigma}
\sigma_{\rho \sigma}$. In fact, $G_2={F_1 / 2}$,
so that we need to compute only the form factor $F_1$
at the kinematical point $q^2=0$.

As usual in the QCD sum rules approach,
we consider the three-point function correlator:
\be T_{\alpha \mu \nu}(p, p^\prime, q) = (i)^2 \int dx \; dy \;
e^{i(p^\prime \cdot x - p \cdot y)}
\bra {0} T (J_\alpha(x) \hat O_{\mu \nu} (0) J^\dagger_5 (y) ) \ket {0}
\hskip 5pt ,\label{eq8} \ee

\noindent
where the currents $J_\alpha$ and $J_5$
have the same quantum numbers of the $K^*$ and $B$ mesons, respectively, i.e.
\bea J_\alpha (x) & = & \bar q(x) \gamma_\alpha s(x) \hskip 5pt,
 \nonumber \\
 J_5 (y) & = & \bar q(y) i \gamma_5 b(y) \hskip 5pt, \label{eq9} \eea

\noindent whereas the operator
$\hat O_{\mu \nu}(0)$ is given by
\be \hat O_{\mu \nu} (0) = \bar s (0) {1\over 2} \sigma_{\mu \nu} b(0)
\hskip 5pt ,\label{eq10} \ee

\noindent ( $\sigma_{\mu \nu}= {i \over 2} [\gamma_\mu, \gamma_\nu]$).
Using the decomposition
\be T_{\alpha \mu \nu} q^\nu = i \epsilon_{\alpha \mu \rho \sigma} p^\rho
p^{\prime \sigma}  T(p^2, p^{\prime 2}, q^2) \label{eq11} \ee

\noindent we can compute the invariant amplitude $T$ in QCD through the
Operator Product Expansion
for $p^2, p^{\prime 2}$ large and spacelike
\footnote{The extrapolation from $q^2$ large and spacelike to $q^2=0$ is
possible due to the large $b$-quark mass.}, obtaining a perturbative
term and non-perturbative power corrections
 parameterized in terms of quark and gluon condensates.
Keeping the terms up to $D=5$, and in the limit $m_s=0$, we have:

\bea T^{QCD} (p^2, p^{\prime 2}, q^2=0) & = & {3 m_b^4 \over 8 \pi^2} \int ds
\; ds^\prime {s^\prime \over (s - s^\prime)^3 }
{ 1 \over (s- p^2)  (s^\prime - p^{\prime 2})}
\nonumber \\
& - & { m_b \over 2} <\bar q q> {1 \over (m^2_b - p^2) (- p^{\prime 2})}
\nonumber \\
& + & {m_b \over 2} g <\bar q \sigma G q> \Big[
{ m_b^2  \over 2 (m^2_b - p^2)^3 (- p^{\prime 2})}
\nonumber \\
& + &
 { m_b^2  \over 3 (m^2_b - p^2)^2 (- p^{\prime 2})^2} -
{ 1 \over 2 (m^2_b - p^2)^2 (- p^{\prime 2})}\Big ] \hskip 5pt.
\label{eq12} \eea
\noindent
In eq.(\ref{eq12}) we have not included the $O(\alpha_s)$ correction in the
perturbative term.
As it will become clear shortly, the one loop perturbative
contribution turns out to be rather small,
on account of the restricted integration range in
the dispersion relations. Hence, it is reasonable to expect the two-loop
contribution to be negligible.
Also, the $D=4$ non-perturbative term from the gluon condensate
is not yet known. However, we do not expect this term to influence
much the results. For instance, in
other QCD sum rule applications to heavy-light quark
systems, where the Wilson coefficient of the gluon condensate has been
calculated, it has been proven to contribute very little to the sum rules.
Current uncertainties in the values of the
leptonic decay constants will dominate the final error on $F_1(0)$.

The invariant amplitude $T (p^2, p^{\prime 2}, q^2)$ can be related,
by means of a
double dispersion relation, to a
 hadronic spectral density which gets contributions
from the lowest lying resonances, the $B$ and $K^*$ mesons, plus higher
resonances and a continuum:
\bea T^H (p^2, p^{\prime 2}, 0) & = & f_B { m_B^2 \over m_b} f_{K^*} m_{K^*}
{ 1 \over (p^2 - m^2_B) (p^{\prime 2} - m^2_{K^*}) } F_1(0)
\nonumber\\
\nonumber \\
& + &
higher\; resonances + continuum.
\label{eq13} \eea

\noindent In eq.(\ref{eq13}), $f_{K^*}$ and $f_{B}$
are the leptonic decay constants of the $K^*$ and
$B$ mesons, respectively, defined by:
$\bra{ 0} J_\mu \ket {K^*(p^\prime, \eta)} = f_{K^*} m_{K^*} \eta_\mu$,
and $\bra{ 0} J_5 \ket {B(p)} = f_B { m^2_B / m_b}$.
According to duality, the higher resonance states and the continuum
 contribution
to the hadronic spectral function can be modelled by perturbative QCD.
A prediction for $F_1(0)$ can then be obtained
by equating the hadronic and the QCD sides of the sum rule
in the  duality region
\bea m_b^2 < & s & < s_0 \nonumber\\
         0 < & s^\prime & < \bar s = min(s-m_b^2, s^\prime_0) \hskip 5pt,
\label{eq13a} \eea
\noindent where $s_0$ and $s^\prime_0$ are effective thresholds.

The convergence of the series involving the power corrections, and
 the dominance of the
lowest lying states, can be improved by performing a double Borel transform,
with the result:
\newpage
\bea
&  f_B &{ m_B^2 \over m_b} f_{K^*}  m_{K^*}
\exp{[-{ m^2_B\over M^2} - { m^2_{K^*}\over M^{\prime 2}}]} \;
 F_1(0) = \nonumber \\
& = & {3 m_b^4 \over 8 \pi^2}
\int_{m_b^2}^{s_0} ds \int_0^{\bar s}
ds^\prime {s^\prime \over (s - s^\prime)^3 }
\exp[-{ s\over M^2} -{ s^\prime \over M^{\prime 2}}]
\nonumber \\
\nonumber\\
& - & { m_b \over 2} <\bar q q> \exp[-{ m^2_b\over M^2}] \;
\Big[  1 - m_0^2 \Big (
{ m_b^2  \over 4 M^4} + { m_b^2  \over 3 M^2 M^{\prime 2}} -
{ 1 \over 2 M^2} \Big ) \Big ]
\hskip 5pt , \label{eq15} \eea
\\
 where we have adopted the notation
$ g <\bar q \sigma G q> = m^2_0 <\bar q q>$.

\noindent
Equation (\ref{eq15}) is the sum rule for $F_1(0)$. We use the following
values of the parameters:
$<\bar q q>=(-0.23\; GeV)^3$, and $m^2_0=0.8 \; GeV^2$;
the effective thresholds are varied in the range $33 - 36 \; GeV^2$ for $s$
and $1.4 - 1.8 \; GeV^2$ for $s^\prime$.
The value of $f_{K^*}$ can be obtained from the decay
$\tau^- \to K^{*-} \nu_\tau$:
$f_{K^*}=0.22\pm0.01 \; GeV$; as for $m_b$ and $f_B$ we use
$m_b=4.6 \;GeV$  and, consequently,
$f_B=0.18 \pm 0.01 \;GeV$ as computed in
\cite{noi}
\footnote{For a recent calculation of the beauty quark
mass see \cite{DOMINGUEZ}. }.

The standard numerical analysis of eq.(\ref{eq15}) consists in
looking for a region in the
$(M^2, M^{\prime2}$) space where
the value of $F_1(0)$ is reasonably independent
of the Borel variables $M^2, M^{\prime2}$, and of the thresholds $s_0$,
$s^\prime_0$. In addition, one may look for
a hierarchy in the series of power
corrections, although there are  a few known examples where this
hierarchy is absent, without affecting the reliability of the results.
It turns out that, in the present case, the perturbative contribution in
(\ref{eq15})
is smaller than the $D=3$
non-perturbative term, due
to the tiny integration region which corresponds
to reasonable choices of the thresholds.
For this reason we are forced to relax
the criterion of the hierarchy among all the terms in the power series
expansion.
 In this way,
 as shown in fig.(1), we find a stability region
in correspondence to the value $F_1(0)=0.35 \pm 0.05$ (the error is
obtained by varying the parameters in their allowed intervals).
This result allows us to estimate the fraction of inclusive
$b \to s \gamma$ decays
represented by the exclusive channel $B \to K^* \gamma$:

\be {\Gamma(B \to K^* \gamma) \over \Gamma(b \to s \gamma)}
= \Big( {m_B \over m_b} \Big)^3 \; \Big(1 - {m^2_{K^*} \over m^2_B} \Big)^3 \;
|F_1(0)|^2 = 0.17 \pm 0.05 \hskip 5pt.\label{eq16} \ee

\noindent
Using the computed value of the inclusive branching ratio we find
$ BR(B \to K^* \gamma) =(4 \pm 1) \times 10^{-5}$, to be compared
with eq.(\ref{eq0}).

The result in eq.(\ref{eq16}) is smaller than that obtained in
\cite{ALIEV} by roughly a factor of 2.
First, there is a technical difference between our determination of
$F_1(0)$ and that of \cite{ALIEV}. While we start from the three-point
function (\ref{eq8}), which leads to a one-step determination of $F_1(0)$
from the sum rule (\ref{eq15}), the authors of \cite{ALIEV} use the
equations of motion to rewrite $F_1(0)$ as a linear
combination of
two form factors which satisfy two different QCD sum rules. We find that
this procedure is unnecessary, and that it may introduce artificial
 uncertainties in
the results for $F_1(0)$. Second, the authors in \cite{ALIEV} use
$f_B \simeq 130 \; MeV$, a value which is not confirmed by recent
lattice and QCD sum rules analyses both for finite and infinite heavy quark
masses \cite{fb,Martinelli}.

A calculation similar to the one described above
can be done  for the transition $B \to K_1 \gamma$.
 Since the electromagnetic penguin is a spin-flip operator,
the $K_1$ meson is the $^3P_1$,
$1^+$ orbital excitation in the kaon system and, in the framework of QCD
sum rules, it can be interpolated by the current
$J_\alpha= \bar q \gamma_\alpha \gamma_5 s$.

As it is well known, there are two $1^+$ states in the strange particle
spectrum, i.e. $K_1(1270)$ and $K_1(1400)$, whose full
widths are: $90 \pm 20 \; MeV$ and $174 \pm 13 \; MeV$, respectively
\cite{PDG}.
The leptonic decay
constants of these states  can be estimated
using experimental information on the hadronic decays of the $\tau$ -lepton.
{}From the measured branching
ratio \cite{Kolanoski}
BR($\tau \to (K_1(1270)+ K_1(1400) ) \; \nu_\tau) \;  = 1.14\pm0.5 \times
10^{-2}$
and using  \cite{Kolanoski} for the fraction of the $\tau$ decay into
$K_1(1270)$ and
$K_1(1400)$, the values:
$0.48 \pm 0.38$ and
$0.66 \pm 0.35$, respectively, we find:
$f_{K_1(1270)}=180\pm 81 \; MeV$ and $f_{K_1(1400)}=258\pm 89 \; MeV$.
These values are in agreement, within  errors,  with
a previous estimate
of the leptonic decay constant of the $^3P_1$ octet \cite{Reinders}
, based on two-point function QCD sum rules.

It is likely that the two observed states are a mixture of the $^1P_1$ and
$^3P_1$ states induced e.g. by the $K^* \pi$ channel.
Since the mixing angle is not well known,
we assume that the $^3P_1$ state is an intermediate state with mass
$1300 \; MeV$ and leptonic decay constant $f_{K_1}=200 \; MeV$.
 Using the threshold
$s^\prime_0=3-4 \; GeV^2$ we obtain the value
$F_1(0)=0.5 \pm 0.1$, which corresponds to
\be {\Gamma(B \to K_1 \gamma) \over \Gamma(b \to s \gamma)}=
0.30 \pm 0.15 \hskip 5pt.\label{eq17} \ee

\noindent
This result indicates that this transition could be dominant with respect
to the
$B \to K^* \gamma $ decay. It is rather amusing to recall that while
both the $^3P_1$ and the $^1P_1$ states are produced in $\tau$-lepton decays,
as well as in hadronic reactions, the latter is forbidden in the radiative
decay of the B meson. Hence, the observation of the above decay channel could
shed light on the mixing between the two states $K_1(1270)$ and $K_1(1400)$.

Our analysis of the $B \to K^* \gamma$ transition can be concluded
by discussing the
limit $m_b \to \infty$. Although this does not affect our estimates of
the exclusive radiative decays, it is important in itself as it can
provide information  on the form factors entering the semileptonic decays
of the B-meson. In fact,
it is known that in the limit of an infinitely heavy $b$ quark
one can relate the hadronic matrix element of the effective operator
responsible for the transition $b \to s \gamma$ to the matrix
elements of the weak currents between
a heavy and a light meson \cite{isgur}, by using the flavour and spin
symmetries of the Heavy Quark Effective Theory (HQET) \cite{gen}.
Indeed, one obtains the relation:
\\
\be
F_1(q^2) =  \left\{\frac{q^2+m_B^2-m_{K^*}^2}{2 m_B}
\frac{V(q^2)}{m_B+m_{K^*}}-\frac{m_B+m_{K^*}}{2 m_B}A_1(q^2)\right\}
\label{eq20} \ee
\\
 among the form
factor $F_1$ responsible for the transition $B \to K^* \gamma$
and the semileptonic form factors $V(q^2)$ and $A_1(q^2)$ defined by:
\bea
\langle K^*(p^\prime,\eta)|\bar s\gamma_{\mu}(1-\gamma_5)b|{\bar B}
(p)\rangle  &=& \frac{2 V(q^2)}{m_B+m_{K^*}}
\epsilon_{\mu\nu\lambda\sigma} \eta^{*\nu} p^\lambda p^{\prime\sigma}
\nonumber\\
\nonumber \\
&+& i (m_B + m_{K^*}) A_1(q^2) \eta^*_{\mu} +...
\label{eq21}\eea
\\
($\bar B= B^-$ or ${\bar B}^0$); in (\ref{eq21})
the ellipses denote terms proportional to $(p + p^{\prime})_\mu$
or $q_\mu$.
In using HQET to relate $F_1$ to $V$ and $A_1$ one meets
the following problem. The relation (\ref{eq20}) is supposed to hold for
$q^2\approx q^2_{max}=(m_B-m_{K^*})^2$, i.e. at zero recoil
where the predictions of HQET are
in general reliable, whereas in
$B \to K^* \gamma$ we need $F_1(q^2)$ at $q^2=0$. Moreover
at $q^2\approx q^2_{max}$  we have the following scaling behaviour
of the form factors:
\bea
F_1(q^2_{max}) & \approx & \sqrt{m_b} \nn \\
V(q^2_{max}) & \approx & \sqrt{m_b} \nn \\
A_1 (q^2_{max}) & \approx & \frac{1}{\sqrt{m_b}}.
\label{eq21a}\eea
Therefore, the contribution from $A_1$ in (\ref{eq20}) should, in
principle, be neglected
in the infinite quark mass limit, as it is next-to-leading.
However, $V(q^2)$ and $A_1(q^2)$ might have a different $q^2$ behaviour
and, hence, one cannot exclude that $A_1(0)$ is comparable
to $V(0)$ in the $m_b \to \infty$ limit. QCD sum rules can shed some light
on this issue.

The limit $m_b \to \infty$ in the sum rule for
$F_1(0)$, eq.(\ref{eq15}), can be taken after defining
new low-energy variables $y$, $y_0$ and $E$ as follows:
\bea
s & = & m_b^2 + 2 m_b y \nn \\
s_0 & = & m_b^2 + 2 m_b y_0 \nn \\
M^2 & = & 2 m_b E \hskip5pt;
\eea
and we recall that the scaling law for $f_B$ is given by:
$f_B = \hat f /{\sqrt{m_b}}$, modulo logs.
Before taking the infinite mass limit we observe that the term arising
from the $D=5$ condensates in eq.(\ref{eq15}) seems to diverge. However,
as discussed in \cite{radyushkin,neubert,colangelo}, this
divergence is only apparent. In fact the contributions from the
$D=3$ and $D=5$ condensates should be resummed since they arise from
non local operators: they are only the first two terms in a
power series of the non local operator in the expansion parameter
$m_0^2$. However, the exact form of the non local operator
is unknown, and the resummation procedure introduces
an ambiguity. In \cite{radyushkin,neubert,colangelo} an exponential
form $\exp(-m_0^2 A)$ was adopted, but a behaviour such as
$(1+\frac{m_0^2 A}{n})^{-n}$ ($n=1,2,...$) is also acceptable since
the power series of both forms starts with the same term: $1-m_0^2 A$.
We observe, though, that for
$B \to K^* \gamma$, and with $m_b \approx 4.6$ GeV ( no doubt a large mass),
eq.(\ref{eq15}) shows the feature that the
non-perturbative contribution
dominates over the perturbative one. Should we resum the non
perturbative contributions in the exponential form, we would obtain
the unlikely result that, when $m_b \to \infty$, the dominant contribution
to the sum rule would be provided by the perturbative term, contrary
to the expectation based on the rather large value $m_b \approx 4.6$ GeV.
Therefore, we shall assume the form
\be
(1+\frac{m_0^2 A}{n})^{-n} \, .\label{eq22}
\ee
Moreover, if one adopts the reasonable criterion that when $m_b \to
\infty$, the quark condensate contribution still dominates the sum rule,
one gets $n=1$ and therefore the sum rule takes the form:
\bea
f_{K^*} m_{K^*} \hat f \exp(- {m^2_{K^*} \over M^{\prime 2} }) F_1 (0) &=&
{3 \over 4 \pi^2  m_b^{3/2} } \int_0^{y_0} dy \int_0^{+ \infty} ds^\prime
s^\prime \; \exp[ - {y \over E} - {s^\prime  \over M^{\prime 2}}] \nonumber \\
&-& {\sqrt m_b \over 2} <\bar q q>
\Big( 1 + {m_0^2 m_b \over 6 E M^{\prime 2}} \Big)^{-1}
\label{eq23}\eea
which shows that $F_1(q^2)$ behaves near the origin as follows:
\be
F_1 (0)  \approx  \frac{1}{\sqrt{m_b}} \hskip 5pt. \label{eq24}
\ee
The consequence is
that the behaviour of $F_1(q^2)$ between
$q^2 \approx 0$, eq.(\ref{eq24}), and $q^2 \approx q^2_{max}$,
 eq.(\ref{eq21a}),
can be interpolated by the simple pole formula suggested by a
dispersion relation dominated by the nearest $1^-$ resonance:
\be
F_1(q^2) = \frac{F_1(0)}{1-{q^2}/{m^2_{B^*_s}}} \hskip 5pt.
\label{eq25}
\ee

Let us now examine the validity of eq.(\ref{eq20}) at $q^2=0$. We can follow
the same method discussed above to derive the behaviour of the
sum rules for $V(0)$ and $A_1(0)$ in the $m_b \to \infty$ limit.
We obtain:
\bea
f_{K^*} m_{K^*} \hat f \exp(- {m^2_{K^*} \over M^{\prime 2} }) V(0) &=&
- {3 \over 4 \pi^2  m_b^{3/2} } \int_0^{y_0} dy \int_0^{+ \infty} ds^\prime
s^\prime \; \exp[ - {y \over E} - {s^\prime  \over M^{\prime 2}}] \nonumber \\
&-& {\sqrt m_b \over 2} <\bar q q>
\Big( 1 + {m_0^2 m_b \over 6 E M^{\prime 2}} \Big)^{-1}
\hskip 5pt ,\label{eq26}\\
\nonumber\\
\nonumber \\
f_{K^*} m_{K^*} \hat f \exp(- {m^2_{K^*} \over M^{\prime 2} }) A_1(0) &=&
-{3 \over 4 \pi^2  m_b^{3/2} } \int_0^{y_0} dy \int_0^{+ \infty} ds^\prime
s^\prime \; \exp[ - {y \over E} - {s^\prime  \over M^{\prime 2}}] \nonumber \\
&+& {\sqrt m_b \over 2} <\bar q q>
\Big( 1 + {m_0^2 m_b \over 6 E M^{\prime 2}} \Big)^{-1}
\hskip 5pt.\label{eq27}
\eea

The behaviour in eqs.(\ref{eq26})-(\ref{eq27}) is
 compatible with the scaling
laws in eq.(\ref{eq21a}) only
if the form factors $V$ and $A_1$ have different $q^2$
dependence, for example  $V$ is a simple pole analogous to
(\ref{eq25}) whereas $A_1$ is nearly constant in $q^2$. This is
in agreement with the observation made in \cite{Ball}, where the $q^2$
dependence of the form factors of the semileptonic $B \to \rho$ transition
has been explicitly investigated.

We notice, in closing, that using eqs.(\ref{eq23},\ref{eq26},
\ref{eq27}), the relation  (\ref{eq20}) is verified at $q^2=0$.
This is a direct confirmation of  the conjecture put forward in \cite{isgur}
and in \cite{Burdman}, that in  heavy-to-light transitions
the heavy quark stays almost on its mass shell, and that the relations
obtained at the zero recoil point  remain applicable over the full $q^2$
 region.
\vskip 1cm
\noindent {\bf Acknowledgments}

\noindent One of us (CAD) wishes to thank Daniele Amati for his kind
hospitality at SISSA, where this work has been done. The work of (CAD)
has been supported in part by the Foundation for Research Development.

\newpage
\vskip 1cm
\begin{center} {\bf Figure captions} \end{center}
\vskip 1cm
\noindent {\bf Fig. 1.} $F_1(0)$ versus the Borel parameters
$M^2, M^{\prime^2}$ for $s_0=36 \; GeV^2$, $s^\prime_0=1.8 \; GeV^2$.
\end{document}